\documentclass[conference]{IEEEtran}
\IEEEoverridecommandlockouts

\usepackage{changepage}
\usepackage{cite}
\usepackage{amsmath,amssymb,amsfonts}
\usepackage[spanish, english]{babel}
\usepackage{booktabs} 
\usepackage{algorithmic}
\usepackage{graphicx}
\usepackage{textcomp}
\usepackage{caption}
\usepackage{float}
\usepackage{longtable}
\usepackage{slashbox}
\usepackage{epstopdf}
\usepackage{epsfig}
\usepackage{graphicx}
\usepackage{tabulary}
\def\BibTeX{{\rm B\kern-.05em{\sc i\kern-.025em b}\kern-.08em
    T\kern-.1667em\lower.7ex\hbox{E}\kern-.125emX}}
    
\newcommand{\figref}[2][]{Fig.\@~\ref{#2}#1}

\IEEEoverridecommandlockouts
\IEEEpubid{\begin{minipage}{\textwidth}\ \\[12pt]
  * Corresponding author
\end{minipage}}

\begin{document}
\small
\title{Experimental Black-box System identification and control of a Torus Cassegrain Telescope}

\author{\IEEEauthorblockN{Xiomara Camacho Medina}
\IEEEauthorblockA{\textit{Escuela de Ciencias Exactas e Ingenier\'ia} \\
\textit{Universidad Sergio Arboleda}\\
Bogot\'a DC, Colombia \\
}
\and
\IEEEauthorblockN{Tatiana Manrique*}
\IEEEauthorblockA{\textit{Escuela de Ingenier\'ia y Ciencias B\'asicas} \\
\textit{Universidad EIA}\\
 km 2+200 variante aeropuerto JMC, 055428\\
 Envigado, Colombia \\
dolly.manrique@eia.edu.co}
}

\maketitle

\begin{abstract}
\noindent The Astronomical Observatory of the Sergio Arboleda University (Bogot\'a, Colombia) has as main instrument a Torus Classic Cassegrain telescope. In order to improve its precision and accuracy in the tracking of celestial objects, the Torus telescope requires to be properly automated. The motors, drivers and sensors of the telescope are modelled as a black-box systems and experimentally identified. Using the identified model several control laws are developed such as Proportional, Integral and Derivative (PID) control and State-feedback control for the velocity and position tracking and disturbance-rejection tasks.
\end{abstract}


\section{Introduction}

\begin{figure}[!h]
\centering
\includegraphics[trim= 30mm 0mm 30mm 0mm, scale=0.32]{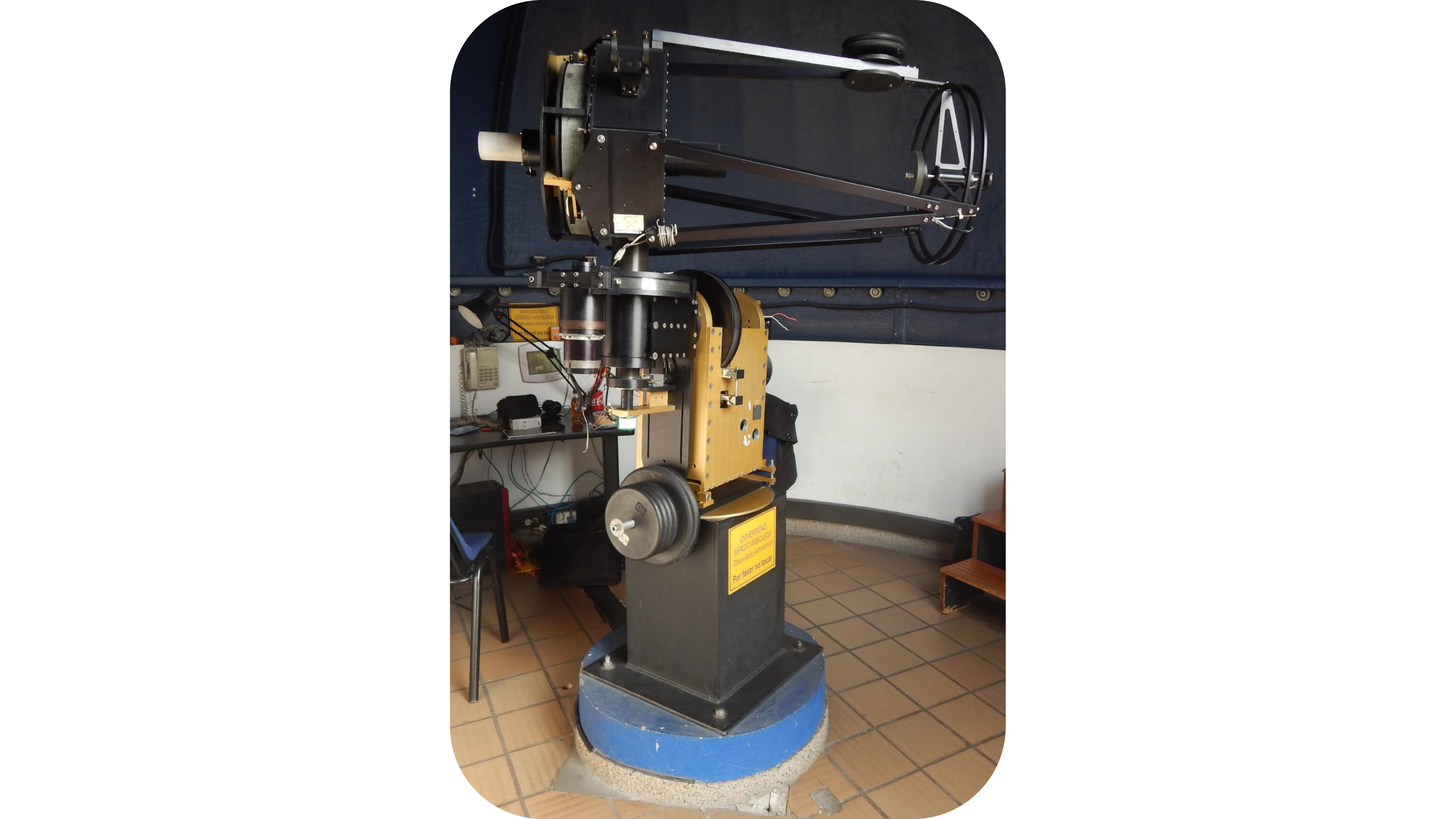}
\captionsetup{justification=centering}
\caption{\small{Torus Classic Cassegrain telescope available at the Astronomical Observatory of the Sergio Arboleda University (Bogot\'a, Colombia).}}
\label{fig:Torus}
\end{figure}

The Astronomical Observatory of the Sergio Arboleda University\footnote{http://www.usergioarboleda.edu.co/escuela-de-ciencias-exactas-e-ingenieria/observatorio-astronomico} (Bogot\'a, Colombia), uses a Torus Classic Cassegrain telescope to track objects in the celestial sphere (see \figref{fig:Torus}). 

\begin{figure}[h]
\centering
\includegraphics[scale=0.3]{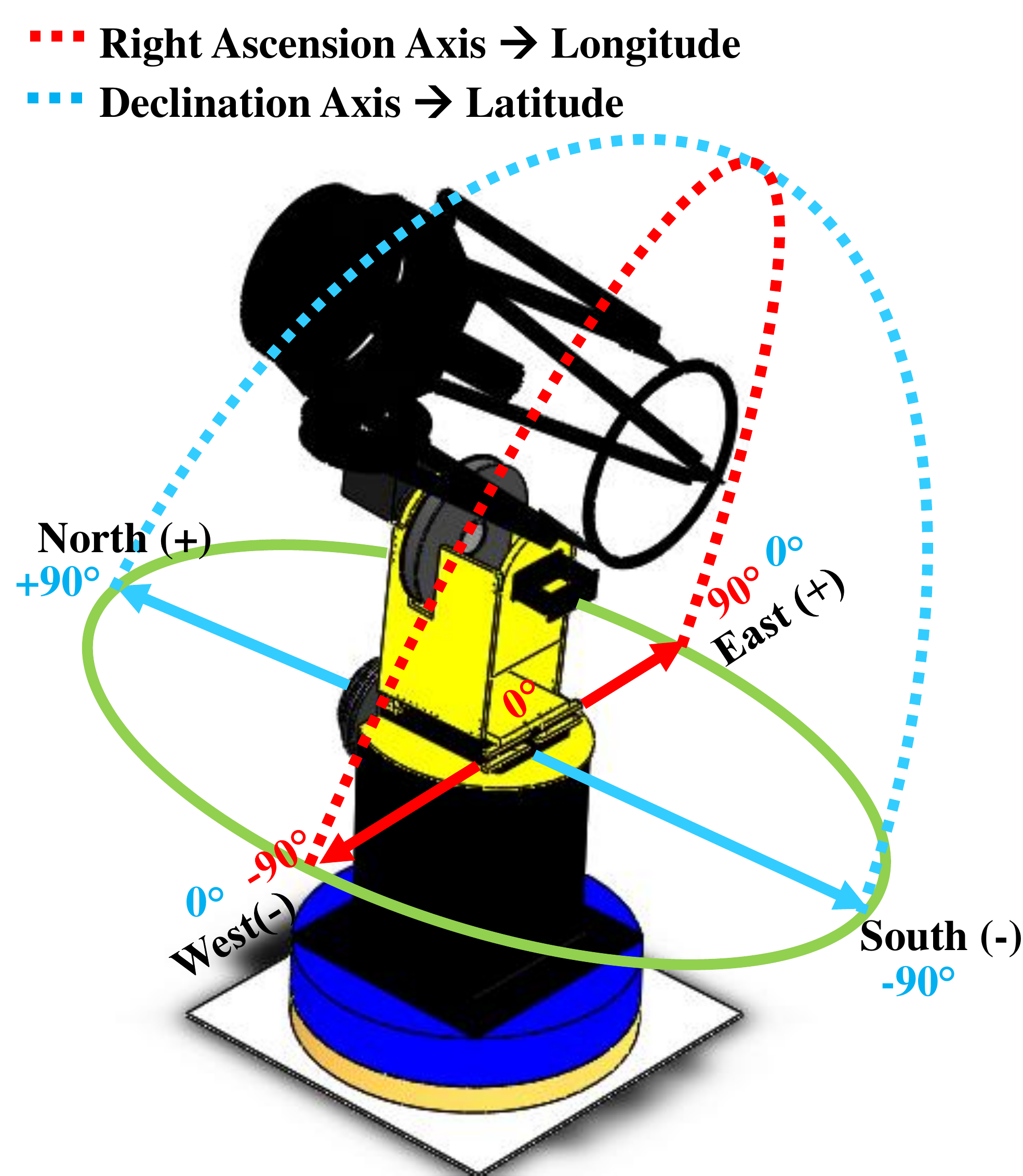}
\captionsetup{justification=centering}
\caption{\small{Torus Cassegrain telescope with German Equatorial Mount (GEM).}}
\label{fig:GEM}
\end{figure}

The Torus Cassegrain telescope designed by Optical Mechanics Inc. \cite{OMI}, and adquired in January 2001 by the Sergio Arboleda University, has a diameter of $0.4m$, a focal length of 4000mm (f/10) and a German Equatorial Mount (GEM) designed especially for the latitude of the geographical location where it is placed. This type of mount in composed by two motors allowing the rotation of the telescope through two axes: straight ascension (or just ascension) and declination  (see \figref{fig:GEM}).

When this telescope was adquired in 2001, it was operated using the Observatory Control and Astronomical Analysis Software (OCAAS)\cite{OCAAS2,OMI} operating system, which allowed its automatic positioning, camera control and focus. Over the years, this software became obsolete and therefore the telescope was not used for a long time. Since the year 2012, new hardware, including new drivers and sensors, has been implemented to acess all the system input and output signals, such as the motors position signals measured with the encoders (position sensors), and  Pulse-width modulation (PWM) signals for the drivers of each motor \cite{TesisAtara}. 

In order to guarantee the precise and accurate tracking of celestial objects, the telescope is now being automated and the proper control strategies are being embedded to control the telescope position and velocity. In the automation process, it is crucial to identify the dynamics of the motors of telescope GEM, so the velocity and position control can be designed and implemented. The position and velocity controllers must satisfy the performance requirements presented in Table \ref{Table1} for both reference tracking and disturbance rejection tasks. Those performance requirements are established taking into account the telescopes parking position: Declination: $-90^{\circ}$ and straight ascension $0^{\circ}$ and the Earth's rotation velocity: $0,004166^{\circ}/s$.

\begin{table}[!h]
 \centering
 \begin{adjustwidth}{-1cm}{0cm}
\begin{tabular}{|c|c|c|c|c|}
\hline 
\textbf{Control Design} &\textbf{Reference Signal}
 & \textbf{$tss \leq$} & \textbf{$\%OS \leq$} &\textbf{$ess \leq$} \\ 
\hline 
Ascension Motor Velocity & $10^{\circ}/s$ & $0,2 s$ & $5\%$ & $0$\\ 
\hline 
Ascension Motor Position & $90^{\circ}$ & $60 s$ & $10\%$ & $0$ \\ 
\hline 
Declination Motor Velocity & $10^{\circ}/ s$ & $0,5 s$ & $10\%$ & $0$ \\ 
\hline 
Declination Motor Position & $180^{\circ}$ &$60 s$ & $10\%$ & $0$ \\ 
\hline 
\end{tabular} 
\end{adjustwidth}
\centering
\caption{\small{Ascension and declination motors velocity and position performance requirements.}}
   \label{Table1}
  \end{table} 

In the present paper the telescope drivers, sensors and motors are modelled as black-boxes which which are experimentally identified. Using the obtained models, PID (Proportional, Integral and Derivative) and State-feedback control laws are sintetized and tested. Also the kinematics calculations (inverse and direct) are developed to determine the effective area in which the telescope can rotate. This paper is organized as follows: In Section \ref{sectionHardware} the main hardware of the Torus telescope, such as sensors and actuators, are presented, Section \ref{sectionExp} develops the System identification of both Ascension and Declination systems from experimental data, in Section \ref{ControlDesign} PID and State-feedback controllers are designed to satisfy performance requirements, in Section \ref{TEST} the performance of each control strategy implemented is recorded, and finally in Section \ref{sectionKinematics} the Kinematics of the Torus telescope is developed and in Section \ref{sectionConclusions} conclusions and perspectives are given.

\section{Torus Cassegrain telescope Hardware}
\label{sectionHardware}
In the present section the main hardware features (actuators and sensors) of the Torus Cassegrain telescope available at the Sergio Arboleda University are described.
\subsection{Gurley 9x20 Encoders}
The Torus telescope has two Gurley 9x20 optical rotary incremental encoders that measures the Straight Ascension and Declination rotation, respectively. The main features of the Gurley 9x20 Torus encoders are \cite{GurleyEncoder}: 
\begin{itemize}
\item Long duration owing to its LED-based illumination (to last at least 100,000 hours).
\item Signal stability guaranteed by differential photo-detectors.
\item Reliability due to its single-board and surface-mount electronics.
\item Resolution of 25400 cycles/rev (101600 counts/rev).
\end{itemize}
\subsection{SSt-1500-R Drivers} 
The Torus telescope has also two SSt-1500-R drivers with high bandwidth and digital vector servo drives for both motors ascension and declination, respectively. Settling time smaller than $1ms$, smooth motion, reliability, universal motor compatibility, and even a neural fuzy logic adaptative control algorithm are the main features of the SSt-1500-R drivers \cite{SST1500}. The drivers inputs are PWM signals sent from the embedded software in the control unit.

\section{Experimental System Identification}
\label{sectionExp}
\begin{figure}[!h]
\centering
\includegraphics[scale=0.2]{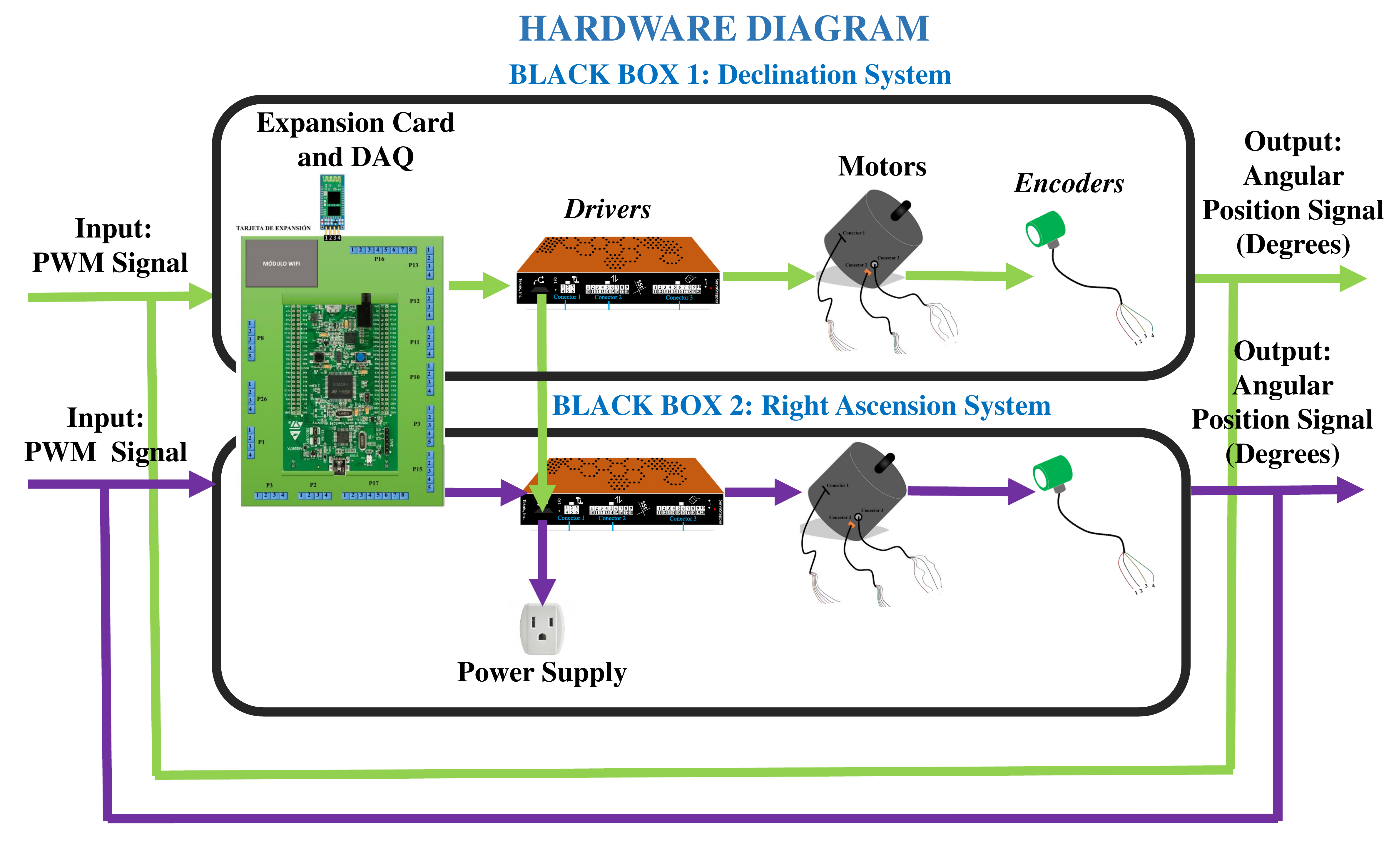}
\captionsetup{justification=centering}
\caption{\small{Ascension and declination black-box systems.}}
\label{fig:Hardware}
\end{figure}

In order to obtain the dynamic models, the ascension and declination systems are modelled as black-boxes that have as input a PWM signal and as output an angular position signal in degrees (see \figref{fig:Hardware}.) 

\begin{figure}[!h]
\centering 
\begin{adjustwidth}{0.5cm}{0cm}
\includegraphics[scale=0.4]{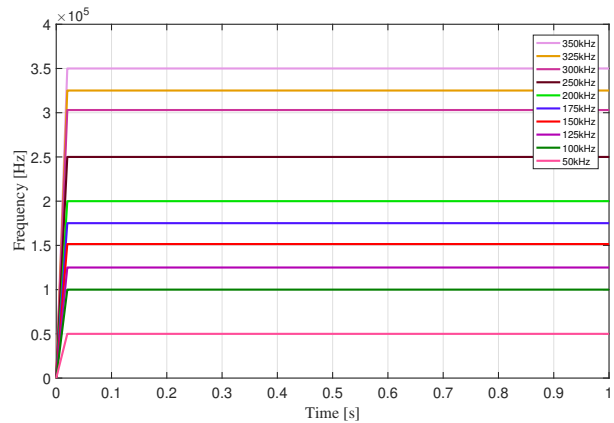}
\caption{\small{Experimental input data set for both Ascension and Declination systems for different PWM input frequencies.}}
\label{fig:InputData}
\end{adjustwidth}
\end{figure}

\begin{figure}[!h]
\centering
\begin{adjustwidth}{-0.4cm}{0cm}
\includegraphics[scale=0.4]{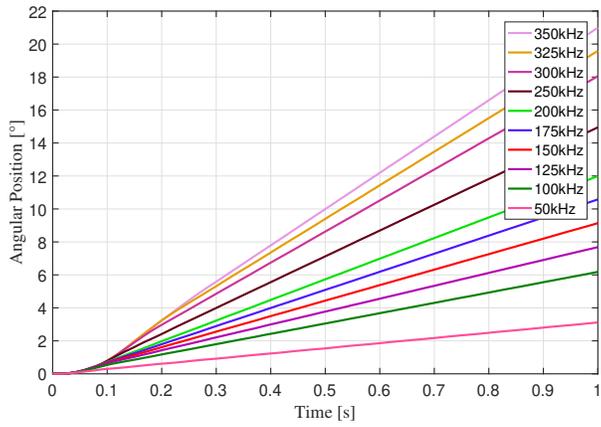}
\captionsetup{justification=centering}
\caption{\small{Experimental position output data set for Ascension system for different PWM input frequencies.}}
\label{fig:PositionData}
\end{adjustwidth}
\end{figure}

\begin{figure}[!h]
\includegraphics[scale=0.4]{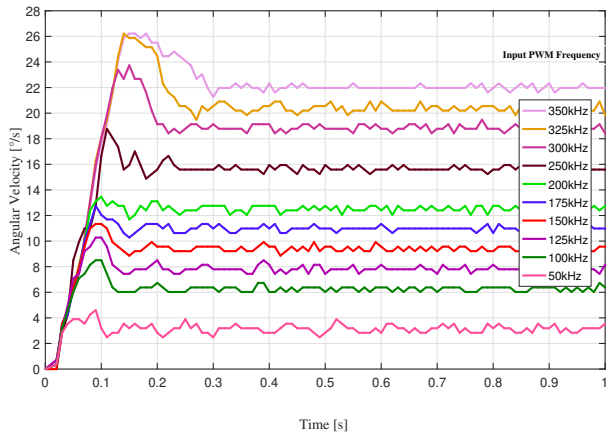}
\captionsetup{justification=centering}
\caption{\small{Experimental velocity output data set for Ascension system for different PWM input frequencies.}}
\label{fig:VelocityData}
\end{figure}

A STM32F4 microcontroller is used to get the position and velocity data of each motor with a $10ms$ sampling time. The maximum frequencies allowed for the PWM signals used as input is $350kHz$. The input and output signals obtained experimentally that will be used as data sets for the systems identification process are depicted in \figref{fig:InputData}, \figref{fig:PositionData} and \figref{fig:VelocityData}. 

From the obtained experimental data it is also possible to conclude that both systems (Straight ascension and Declination) are SISO (Single Input Single Output), both systems are LTI (Linear Time Invariant), the dynamics of the motors position is not BIBO-stable, the velocity behaviour of both systems is a second order with a damping factor $0.2 \leq \zeta \leq 0.7$ and the variation of the duty cycle of the PWM input signal does not affect the outputs, only does the frequency variation. Those analysis performed from experimental data are particularly useful since they allow to properly choose the model structure to identify \cite{Ljung1999}. The transfer function structure is chosen. 

\begin{figure}[!h]
\centering
\includegraphics[scale=0.4]{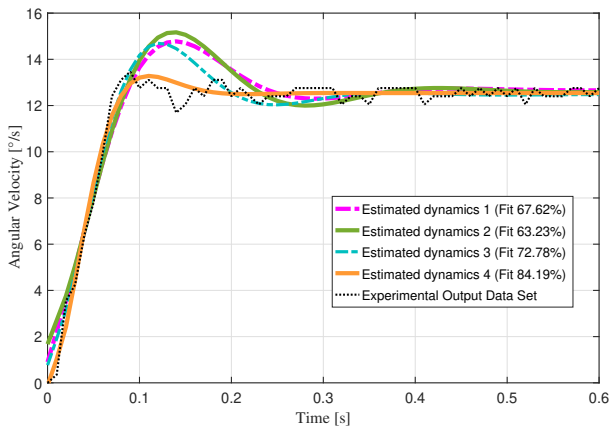}
\captionsetup{justification=centering}
\caption{\small{System identification results for Straight Ascension system velocity dynamics.}}
\label{fig:Validation}
\end{figure}

\begin{figure}[!h]
\centering
\includegraphics[scale=0.3]{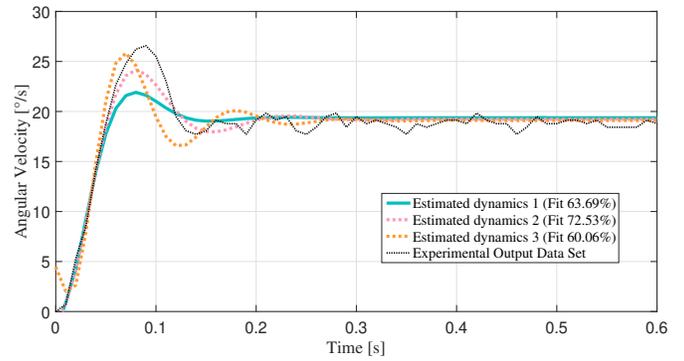}
\captionsetup{justification=centering}
\caption{\small{System identification results for Declination system velocity dynamics.}}
\label{fig:Validation2}
\end{figure}

The velocity dynamics of each motor is identified by using the input data set with the best consistency performance according to the Fit to estimation data (best behaviour at high values) FED, Final prediction error FPE and the Mean-Square error (best behaviour at low values) MSE. The input data sets chosen to perform the system identification are shown in Table \ref{Table2}. The system identification is performed by using the MATLAB System Identification Toolbox \cite{MathWorks2017}. In \figref{fig:Validation} and \figref{fig:Validation2} the best fits to the experimental data are depicted for the Ascension and Declination system dynamics. In Table \ref{Table3} the identified dynamics are presented. 

\begin{table}[!h]
 \centering
\begin{tabular}{|c|c|c|c|c|}
\hline 
\textbf{System}&\textbf{Input data set} &\textbf{FED}
 & \textbf{FPE} & \textbf{MSE} \\ 
\hline 
Ascension & Frequency 250kHz & $76.51\%$ & $0,1062$ & $0.105$ \\ 
\hline 
Declination & Frequency 325kHz & $66.43\%$ & $0,3356$ & $0.3315$ \\ 
\hline 
\end{tabular} 
\caption{\small{Input data sets chosen for system identification.}}
   \label{Table2}
  \end{table} 

\begin{table}[!h]
 \centering
\begin{tabular}{|c|c|c|}
\hline 
\textbf{System}&\textbf{Velocity dynamics} &\textbf{Position dynamics}\\ 
\hline
 & & \\ 
Ascension & $G(s)=\frac{0.09809}{s^{2}+52s+1566.5}$& $G(s)=\frac{0.09809}{s^{3}+52s^{2}+1566.5s}$\\ 
\hline
 & & \\ 
Declination &  $G(s)=\frac{0.1267}{s^{2}+34.72s+2018}$ &$G(s)=\frac{0.1267}{s^{3}+34.72s^{2}+2018s}$\\ 
\hline
\end{tabular} 
\caption{\small{Identified Velocity and Position dynamics.}}
   \label{Table3}
  \end{table}  

\section{Position and velocity control laws}
\label{ControlDesign}
In the present section, the PID and State-feedback controllers are designed. These controllers were selected since they fulfil the required performance characteristics in terms of ess (Steady-state error) equal to $0$ through an integral action, the tss (steady-state time) through a proportional action and derivative action for handling of the \%OS (Overshoot Percentage).The State-feedback allows to include the available information of both states: position and velocity in the control loop, even if they are not simultaneously outputs. From the experimental data and the identified dynamics, the following features can be highlighted for both Ascension and Declination systems:

\begin{itemize}
\item Velocity dynamics is a type $0$ dynamics
\item Position dynamics is a type $1$ dynamics 
\item Velocity dynamics is a BIBO-stable dynamics
\item Position dynamics is marginally stable
\end{itemize}

\begin{figure}[!h]
\centering
\begin{adjustwidth}{-0.8cm}{0cm}
\includegraphics[scale=0.3]{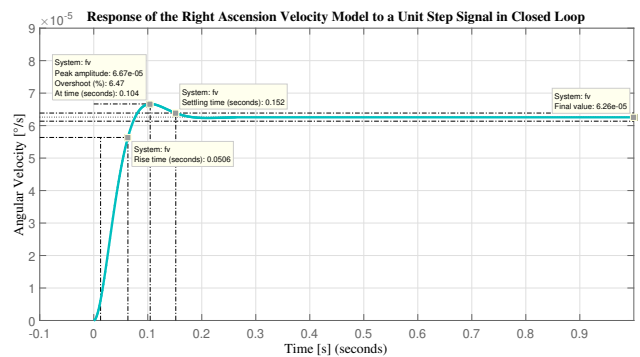}
\captionsetup{justification=centering}
\caption{\small{Ascension velocity response to a unit step input signal.}}
\label{fig:ClosedLoopVel}
\end{adjustwidth}
\end{figure}

\begin{figure}[!h]
\centering
\begin{adjustwidth}{-0.8cm}{0cm}
\includegraphics[scale=0.3]{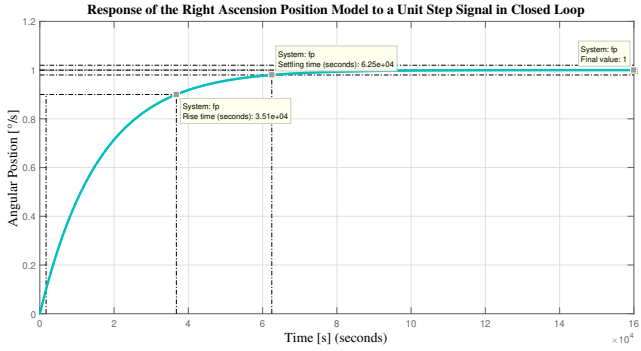}
\captionsetup{justification=centering}
\caption{\small{Ascension position response to a unit step input signal.}}
\label{fig:ClosedLoopPos}
\end{adjustwidth}
\end{figure}

Therefore, when closing the loops with a unitary feedback and no additional control law, the velocity models (Type $0$) do not follow a step reference signal meanwhile the position models do (Type $1$), as shown in \figref{fig:ClosedLoopVel} and \figref{fig:ClosedLoopPos}.

\subsection{PID Controllers Design}

\begin{figure}[!h]
\centering
\begin{adjustwidth}{-0.5cm}{0cm}
\includegraphics[scale=0.4]{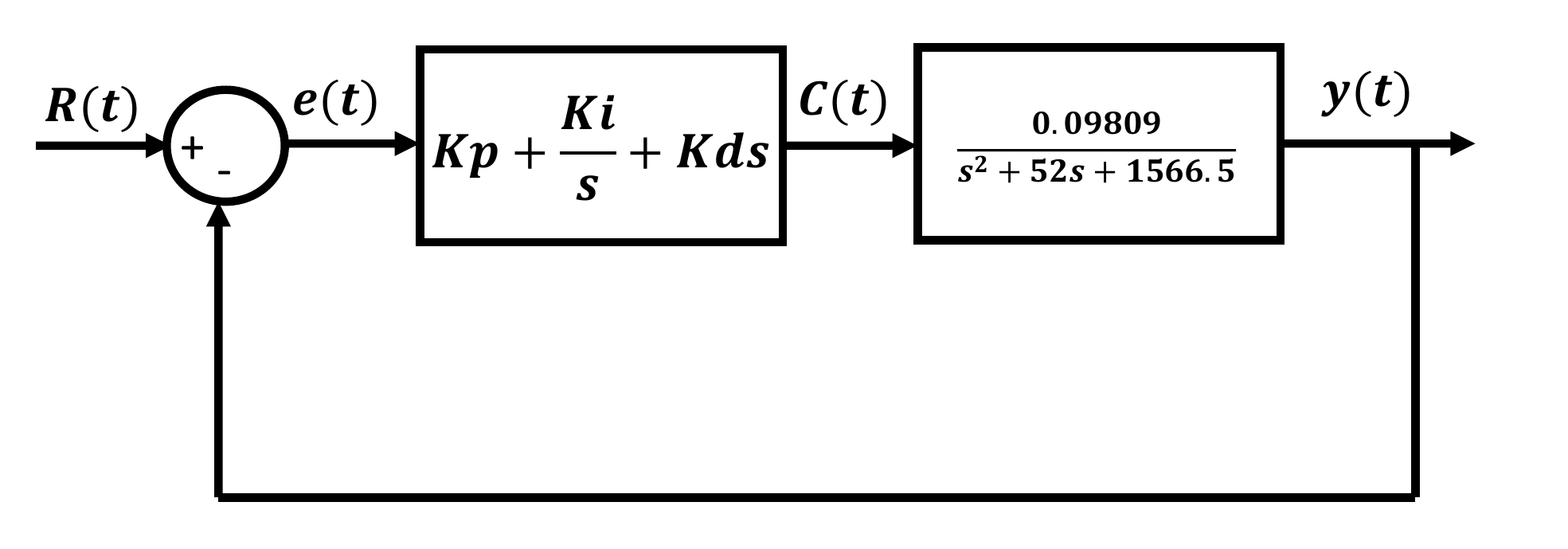}
\captionsetup{justification=centering}
\caption{\small{PID Parallel Configuration.}}
\label{fig:PID}
\end{adjustwidth}
\end{figure}

Despite the previous analysis, PID actions are selected to the velocity controls and PI to the position models (see \figref{fig:PID}), since the input are not guaranteed to be step inputs exclusively and nevertheless ess=0 must be assured. The PID gains calculated to satisfy the performance in Table \ref{Table1} for the models of each system read Table \ref{Table4}. 

\begin{table}[!h]
 \centering
 \begin{adjustwidth}{0cm}{0cm}
\begin{tabular}{|c|c|c|c|}
\hline 
\textbf{Control System}&\textbf{Kp} &\textbf{Ki}&\textbf{Kd}\\ 
\hline 
Ascension Velocity& $73906.004$ &$1664184,219$ & $1916,48$ \\ 
\hline
Ascension Postion & $4277.848$ &$151,872$ & X \\
\hline
Declination Velocity& $-4430.4656$ &$115298,0189$ & $481,80$ \\ 
\hline
Declination Postion& $4530.099$ &$152,818$ & X \\
\hline 
\end{tabular} 
\end{adjustwidth}
\caption{\small{Design of the PID Controllers for the systems.}}
   \label{Table4}
  \end{table}  

The PID and PI controllers are implemented with \textit{Anti Wind-up Filters} to prevent integration wind-up due to the actuator saturation at 350kHz.

\subsection{State Feedback Controllers Design}

\begin{figure}[!h]
\centering
\includegraphics[scale=0.25]{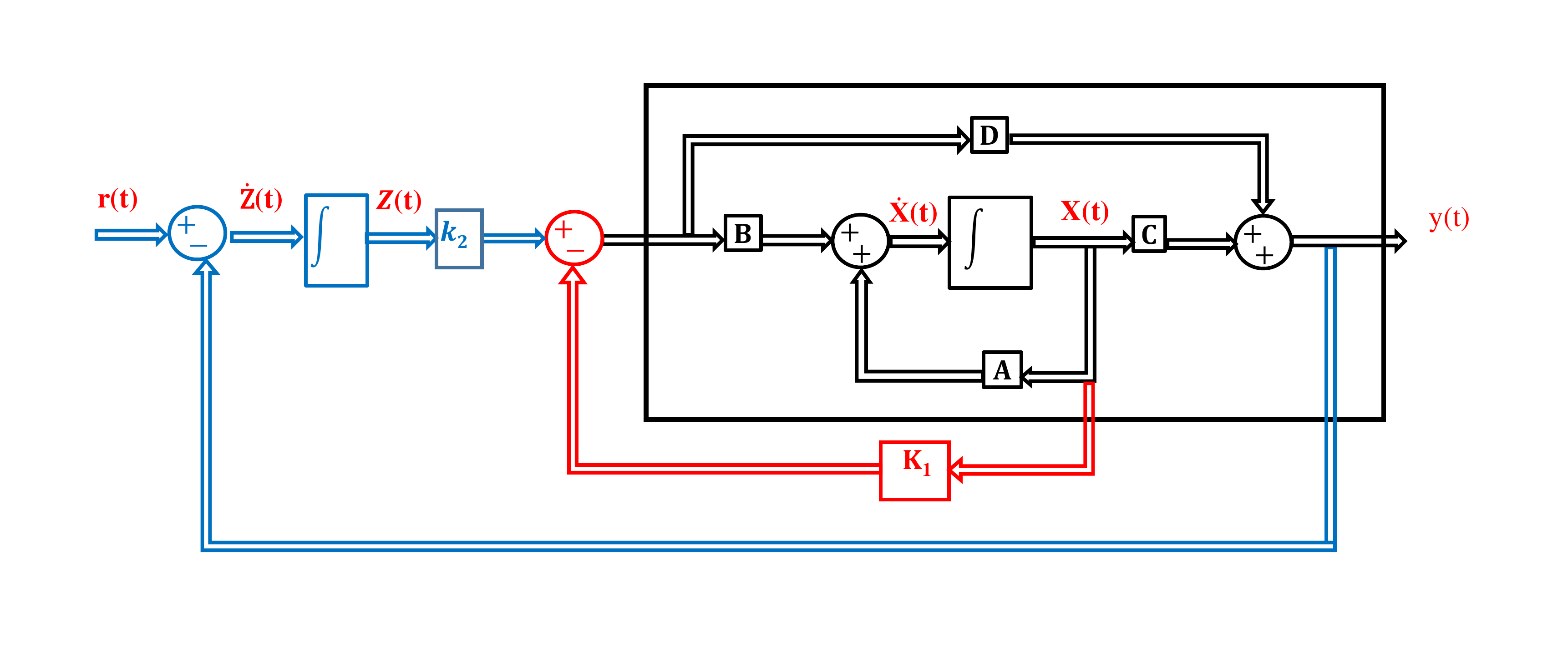}
\captionsetup{justification=centering}
\caption{State Feedback Configuration with a integral gain $k2$}
\label{fig:SF}
\end{figure}

The State-Feedback controllers are designed using the configuration shown in \figref{fig:SF} to guarantee ess=0.

The state-feedback gains found for the control and position control are showed in Table \ref{Table5}, where $K1=[a \; b]$ for velocity dynamics, and  $K1=[a \; b \; c]$ for position models. The parameter $k2$, is the integral constant.

\begin{table}[!h]
 \centering
\begin{tabular}{|c|c|c|}
\hline 
\textbf{Control System}&\textbf{K1} &\textbf{K2}\\ 
\hline 
Asc. Velocity& $a=187,98$ &$1664014,57$ \\ 
&  $b=7248,94$ &\\
\hline
Asc. Position & $a=-45,34$  &$80$\\
&$b=-1542,6$&\\
&$c=83,47$&\\
\hline
Decl. Velocity& $a=61,05$ &$115297,94$ \\ 
&$b=-561,74$&\\
\hline
Decl. Position& $a=-28,05$,&$60$ \\
&$b=-1994,56$&\\
&$c=83,74$&\\
\hline 
\end{tabular} 
\caption{\small{State-Feedback Controllers.}}
   \label{Table5}
  \end{table}  

\section{Analysis of results PID vs. State Feedback Controllers}
\label{TEST}
In the following section trajectory tracking and disturbance rejection tests are performed. The control signals behaviour also analysed to know if the saturations of the system actuator are exceeded.

\subsection{Trajectory Tracking tests}
\begin{figure}[!h]
\centering
\includegraphics[scale=0.27]{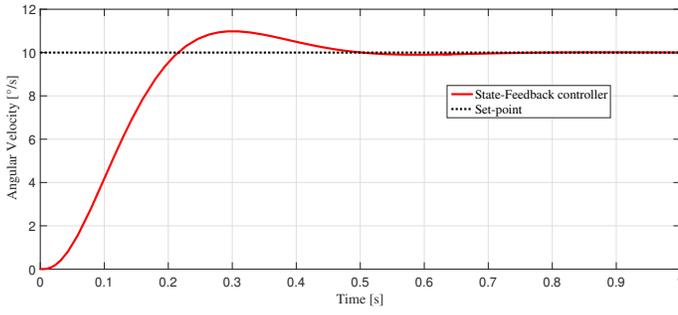}
\captionsetup{justification=centering}
\caption{\small{Velocity set-point tracking response for Declination system whit State-feedback controller.}}
\label{fig:SFDeclination}
\end{figure}

The aim of these tests is to evaluate if the control strategies meet the performance characteristics. Table \ref{Table6} is obtained. In \figref{fig:SFDeclination} the response of the State-feedback controller is depicted when controlling the velocity of declination system.
  \begin{table}[!h]
 \centering
 
\begin{tabular}{|c|c|c|c|c|}
\hline
\multicolumn{2}{|c|}{\backslashbox{\textbf{Controllers}}{\textbf{Performance Characteristics}}}&\textbf{tss} &\textbf{\%OS}&\textbf{ess}\\ 
\hline
&Ascension Velocity & $0.2s$ & $5\%$ & 0 \\
\cmidrule(l){2-5}
\textbf{Performance} &Ascension Position & $60s$ & $10\%$ &0 \\
\cmidrule(l){2-5}
\textbf{requirements}&Declination Velocity &$0.5s$ & $10\%$ & 0 \\
\cmidrule(l){2-5}
&Declination Position & $60s$ & $10\%$ &0 \\
\hline
&Ascension Velocity &$0.2s$ & $10\%$ &0 \\
\cmidrule(l){2-5}
\textbf{PID} &Ascension Position & $60s$& $4,74\%$& 0 \\
\cmidrule(l){2-5}
&Declination Velocity &$0.56s$ & $11.11\%$& 0 \\
\cmidrule(l){2-5}
&Declination Position &$60s$ & $1.62\%$ & 0 \\
\hline
&Ascension Velocity &$0.22s$ &$4.5\%$&0 \\
\cmidrule(l){2-5}
\textbf{State-Feedback} &Ascension Position & $60s$ & $0\%$ &0 \\
\cmidrule(l){2-5}
&Declination Velocity &$0.5s$ & $9.8\%$ & 0 \\
\cmidrule(l){2-5}
&Declination Position & $60s$ & $0\%$ & 0 \\
\hline
\end{tabular} 
  \caption{\small{Results for trajectory tracking tests.}}
 \label{Table6}
  \end{table} 
  
 \subsection{Disturbance Rejection tests}
\begin{figure}[!h]
\centering
\includegraphics[scale=0.5]{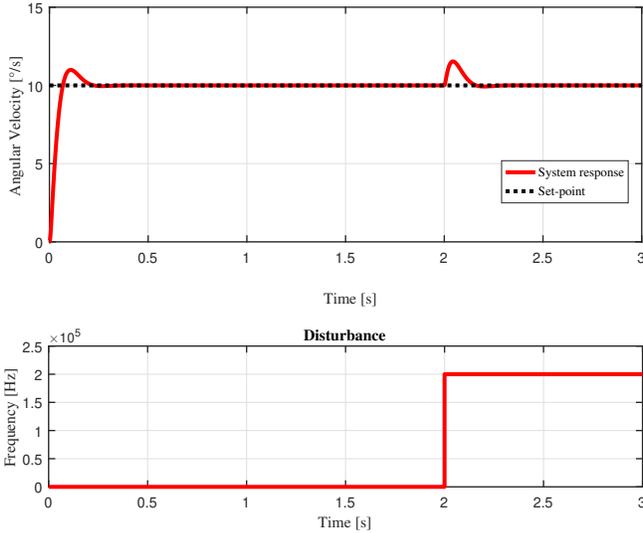}
\captionsetup{justification=centering}
\caption{\small{Velocity disturbance-rejection response for Ascension system with PID controller.}}
\label{fig:PIDDisturbance}
\end{figure}

Disturbance rejection tests seek to evaluate the systems behavior when a perturbation occurs. Table \ref{Table7} collects the controllers performance when disturbance occurs. In \figref{fig:PIDDisturbance} it is shown how the disturbance is applied in the velocity control of the Ascension system when the PID controller is implemented.
 
 \begin{table}[!h]
 \centering
\begin{tabular}{|c|c|c|c|c|}
\hline
\multicolumn{2}{|c|}{\backslashbox{\textbf{Controllers}}{\textbf{Performance Characteristics}}}&\textbf{tss} &\textbf{\%OS}&\textbf{ess}\\ 
\hline
&Ascension Velocity & $0.2s$ & $5\%$ & 0 \\
\cmidrule(l){2-5}
\textbf{Performance} &Ascension Position & $60s$ & $10\%$ &0 \\
\cmidrule(l){2-5}
\textbf{requirements}&Declination Velocity &$0.5s$ & $10\%$ & 0 \\
\cmidrule(l){2-5}
&Declination Position & $60s$ & $10\%$ &0 \\
\hline
&Ascension Velocity &$0.160s$ & $15.3\%$ &0 \\
\cmidrule(l){2-5}
\textbf{PID} &Ascension Position & $60s$& $10\%$& 0 \\
\cmidrule(l){2-5}
&Declination Velocity&$0.5s$ & $115.6\%$& 0 \\
\cmidrule(l){2-5}
&Declination Position &$69.7s$ & $20.11\%$ & 0 \\
\hline
&Ascension Velocity &$0.160s$ &$15.6\%$&0 \\
\cmidrule(l){2-5}
\textbf{State-Feedback} &Ascension Position & $55s$ & $111.11\%$ &0 \\
\cmidrule(l){2-5}
&Declination Velocity &$0.5s$ & $116.4\%$ & 0 \\
\cmidrule(l){2-5}
&Declination Position & $57s$ & $137.77\%$ & 0 \\
\hline
\end{tabular} 
  \caption{\small{Results for disturbance rejection tests.}}
  \label{Table7}
  \end{table} 

\subsection{Control Signals: Saturations}
\begin{figure}[!h]
\centering
\includegraphics[scale=0.4]{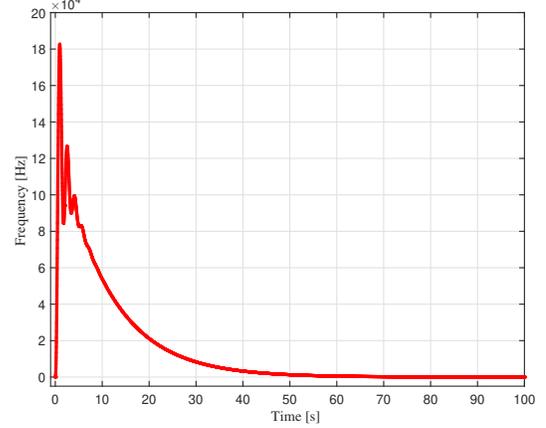}
\captionsetup{justification=centering}
\caption{\small{Control Signal Behaviour Example.}}
\label{fig:ControlSF}
\end{figure}

In Table \ref{Table8} the maximum values achieved by the control signals while performing the above experiments are recorded. The saturation of the actuator is 350kHz. Table \ref{Table8} shows the relevance of implementing Anti-windup filters in PID controllers. In \figref{fig:ControlSF} is depicted the State-feedback control signal obtained while performing the set-point tracking for the Declination position system. 
 \small
  \begin{table}[!h]
 \centering
 \begin{adjustwidth}{0.5cm}{0cm}
\begin{tabular}{|c|c|c|}
\hline
\multicolumn{2}{|c|}{\textbf{Controllers}}&\textbf{Max. PWM Frequency}\\ 
\hline
&Ascension Velocity & $350 kHz$ \\
\cmidrule(l){2-3}
\textbf{PID(with \textit{Antiwind-up})}&Ascension Position & $350 kHz$\\
\cmidrule(l){2-3}
&Declination Velocity & $350 kHz$ \\
\cmidrule(l){2-3}
&Declination Position & $350 kHz$  \\
\hline
&Ascension Velocity & $162 kHz$ \\
\cmidrule(l){2-3}
\textbf{State-Feedback} &Ascension Position & $190 kHz$ \\
\cmidrule(l){2-3}
&Declination Velocity & $170 kHz$ \\
\cmidrule(l){2-3}
&Declination Position &  $180 kHz$ \\
\hline
\end{tabular} 
  \caption{\small{Maximum control signals values.}}
  \label{Table8}
  \end{adjustwidth}
  \end{table}

\section{Torus telescope Kinematics}
\label{sectionKinematics}

\begin{figure}[!h]
\centering
\includegraphics[trim = 0cm 0cm 0mm 0mm,clip,scale=0.2]{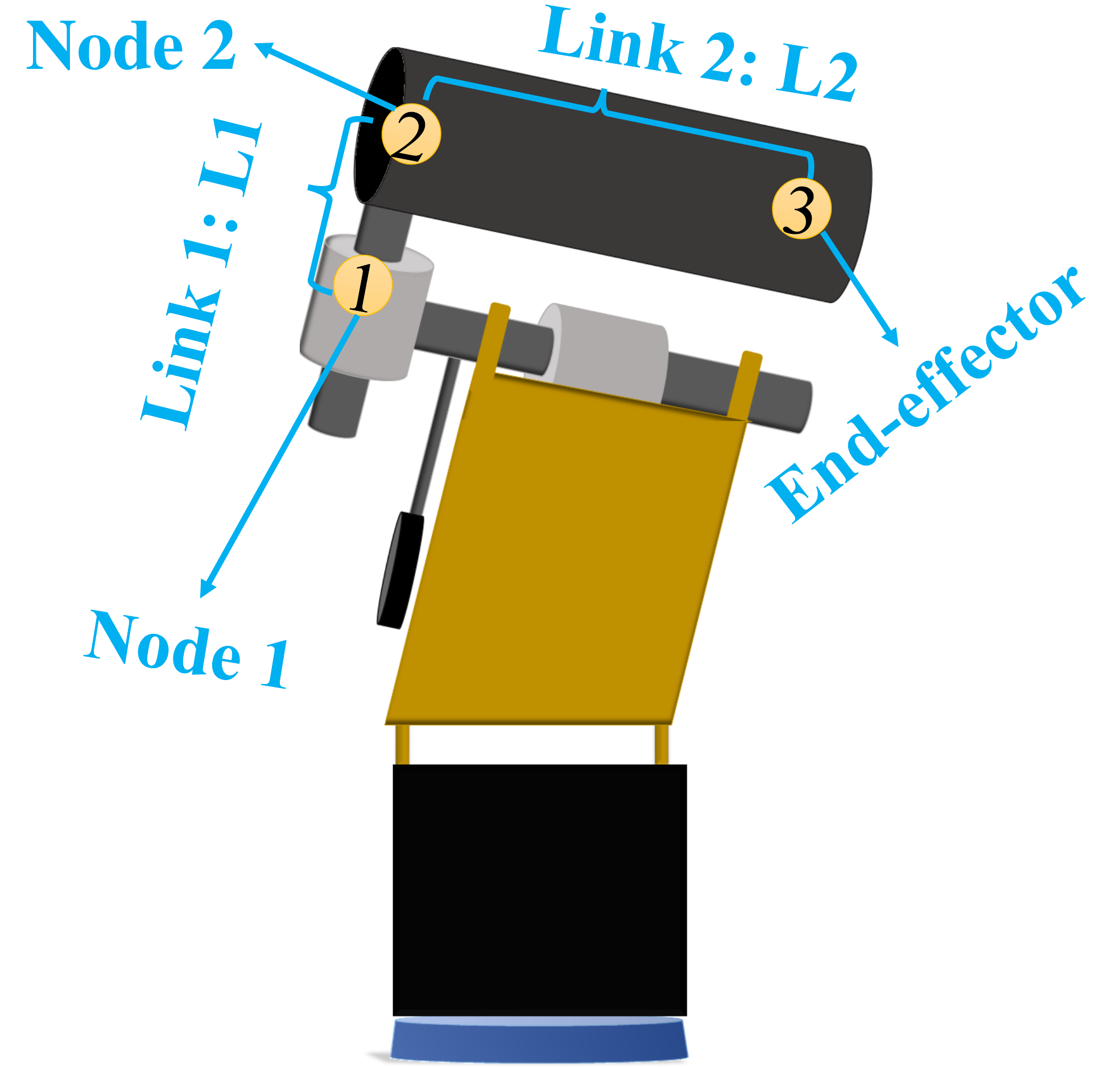}
\captionsetup{justification=centering}
\caption{\small{Links and nodes of Torus Telescope.}}
\label{linksandnodes}
\end{figure}

The inverse and direct kinematics calculation are developed to establish the telescope effective working area. From \figref{linksandnodes} direct and inverse kinematics are obtained. 

\begin{itemize}
\item \textbf{Direct Kinematics:} indicates the final effector (X,Y,Z) position in terms of the nodes position.

\begin{equation}
\begin{split}
Y=\sin(\alpha)L_{2}\cos(\theta_{2}) + (L_{1}\sin(\theta_{1})\\- L_{2}\sin(\theta_{2})\cos(\theta_{1})) \cos(\alpha),
\end{split}
\end{equation}
\begin{equation}
X=L_{1}\cos(\theta_{1}) +L_{2}\sin(\theta_{2})\sin(\theta_{1}),
\end{equation}

\begin{equation}
\begin{split}
Z=\cos(\alpha) L_{2}\cos(\theta_{2})  - (L_{1}\sin(\theta_{1})- \\ L_{2}\sin(\theta_{2})\cos(\theta_{1})) \sin(\alpha).
\end{split}
\end{equation}

\item \textbf{Inverse Kinematic:} indicates the nodes position according to the XYZ position of the final effector. 

\begin{equation}
\sin(\theta_{2})=\sqrt{1- \left( \frac{\sin(\alpha) Y +\cos(\alpha) Z}{L_{2}} \right)^{2} },
\end{equation}

\begin{equation}
\begin{split}
\theta_{1}= \arctan\left( \frac{ Y- L_{2}  \sin(\alpha) \left( \frac{\sin(\alpha) Y+ \cos(\alpha) Z}{L_{2}}\right)}{X\cos(\alpha)}\right) \\ + \arctan \left( \frac{\sqrt{L_{2}^{2}- \left( \sin(\alpha) Y + \cos(\alpha) Z\right)^{2}}}{L_{1}}  \right). 
\end{split}
\end{equation}
\end{itemize}

With these equations, it is possible to simulate the effective area where the telescope operates: 

\begin{figure}[!h]
\centering
\includegraphics[scale=0.5]{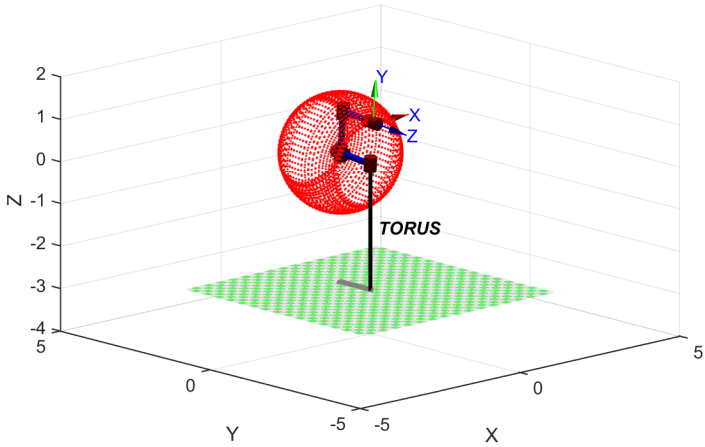}
\captionsetup{justification=centering}
\caption{\small{Effective working area for the Torus Telescope.}}
\end{figure}

\section{Conclusions}
\label{sectionConclusions}
The main conclusions of the present work are: 

\subsection{Experimental System Identification}
\begin{itemize}

\item Data acquisition of the Declination system is noiser since declination rotation has more friction than ascension rotation.

\item To experimentally identify the position dynamics, is necessary to acquire position data in closed loop, since it exhibits unstable behaviour in open loop.

\item The data noise is more important in low angular velocities. Therefore, only frequencies greater than 50kHz were taken into account for the experimental identification.  

\item Data noise filtering was first used in frequencies below 50kHz, nevertheless, this idea was discarded because the transient information was significantly delayed.  

\end{itemize}

\subsection{Position and velocity control law} 

\begin{itemize}

\item The PID control guarantees tss but not \%OS due to the saturation in the actuators and despite the anti-windup filter. This indicates that the desired performance is not feasible for those constraints, therefore is necessary to implement non-conventional control laws such as MPC (Model Predictive Control) that takes the system constraints into account.

\item The disturbance Rejection tests shows better performance in PID Controller. This was due to the implementation of anti-windup filter that reduces the integral error.  

\item State feedback Controller shows better results in meeting the general performance requirements, since this control strategy has an additional information: state variables. 

\item \%OS is sacrificed in order to obtain better results in tss in a trade-off to get feasible control problems.

\item PID velocity control of Declination system shows negative velocity signals due to the presence of negative gains in the controllers design. To avoid negative gains different PID configuration must be chosen. 

\end{itemize}

\subsection{Torus Telescope Kinematics}
\noindent The definition of the effective area is very useful to choose positions references that can be actually reached by the telescope.

\begin{adjustwidth}{0.7cm}{0cm}
\tiny
\bibliographystyle{IEEEtran}
\bibliography{article}

\begin{thebibliography}{1}
\providecommand{\url}[1]{#1}
\csname url@samestyle\endcsname
\providecommand{\newblock}{\relax}
\providecommand{\bibinfo}[2]{#2}
\providecommand{\BIBentrySTDinterwordspacing}{\spaceskip=0pt\relax}
\providecommand{\BIBentryALTinterwordstretchfactor}{4}
\providecommand{\BIBentryALTinterwordspacing}{\spaceskip=\fontdimen2\font plus
\BIBentryALTinterwordstretchfactor\fontdimen3\font minus
  \fontdimen4\font\relax}
\providecommand{\BIBforeignlanguage}[2]{{%
\expandafter\ifx\csname l@#1\endcsname\relax
\typeout{** WARNING: IEEEtran.bst: No hyphenation pattern has been}%
\typeout{** loaded for the language `#1'. Using the pattern for}%
\typeout{** the default language instead.}%
\else
\language=\csname l@#1\endcsname
\fi
#2}}
\providecommand{\BIBdecl}{\relax}
\BIBdecl

\bibitem{OMI}
I.~O. Mechanics, ``Ocaas,'' url\;{http://www.opticalmechanics.com/}, Estados
  Unidos, 2016.

\bibitem{OCAAS2}
OCAAS, ``{Observatory Control and Astronomical Analysis System},''
  url\;{vega.inp.nsk.su/~inest/OCAAS/ocaas11-16.ps}, 1998.

\bibitem{TesisAtara}
F.~A.~A. Monta{\~{n}}ez, ``{Sistema electr{\'{o}}nico inal{\'{a}}mbrico para
  comandar la orientaci{\'{o}}n y enfoque del telescopio TORUS de la
  Universidad Sergio Arboleda},'' \emph{Journal of Chemical Information and
  Modeling}, vol.~53, no.~9, pp. 1689--1699, 2013.

\bibitem{GurleyEncoder}
U.~S.~A. Ny, ``{Gurley Series 9x20 Rotary Incremental Encoder},'' no. 800,
  1994.

\bibitem{SST1500}
I.~TEKNIC, ``{Teknic System Manual SST 1500 SERVO SYSTEM},'' 2003.

\bibitem{Ljung1999}
L.~Ljung, \emph{{System Identification}}, segunda ed~ed., P.~Hall, Ed.,
  Londres, 1999.

\bibitem{MathWorks2017}
Mathworks, ``System identification toolbox,''
  url\;{https://www.mathworks.com/help/ident/index.html}, Estados Unidos, 2017.

\end{thebibliography}
\end{adjustwidth}
 \end{document}